\documentclass[
reprint,
superscriptaddress,
amsmath,amssymb,
aps,
pre,
]{revtex4-2}

\usepackage{graphicx}
\usepackage{dcolumn}
\usepackage{bm}
\usepackage[hidelinks]{hyperref}

\PassOptionsToPackage{main=english}{babel}
\usepackage{babel}

\makeatletter
\@namedef{bbl@en}{english}
\@namedef{l@en}{\l@english}
\makeatother

\usepackage{acro}
\usepackage{placeins}

\DeclareAcronym{PS}{
  short = PS,
  long  = polystyrene
}

\begin{document}


\title{It Takes Two to Tribo: Stochastic Charge Evolution in Repeated Binary Collisions of Acoustically Levitated Particles}

\author{Tom F. O'Hara}
\email{tom.ohara@bristol.ac.uk}
\affiliation{School of Civil, Aerospace and Design Engineering, University of Bristol, United Kingdom}

\author{Adrian C. Barnes}
\affiliation{School of Physics, University of Bristol, United Kingdom}

\author{Nathan Croll Dawes}
\affiliation{School of Chemistry, University of Bristol, United Kingdom}

\author{Karen L. Aplin}
\affiliation{School of Civil, Aerospace and Design Engineering, University of Bristol, United Kingdom}

\date{\today}

\begin{abstract}
The mechanisms governing triboelectric charging between insulating particles of the same material remain an open question in nonequilibrium physics, with several competing models proposed to explain observed charging behaviour. Here, we investigate charge evolution in a minimal system consisting of two acoustically levitated particles undergoing repeated binary collisions. To quantify particle charge transfer, purpose-built Faraday cage picoammeters were developed and calibrated. The MultiLev acoustic levitation system was used alongside Ultraino simulations to generate transducer control signals, enabling controlled collision and separation of polystyrene (PS) particles. Although individual collision events exhibit stochastic charge transfer, described by skew-normal distributions, we demonstrate for the first time that the cumulative charge evolution of an individual acoustically levitated particle pair follows the saturation behaviour predicted by the condenser model of triboelectric charging, with fits achieving $R^2 > 0.99$. Under identical conditions, conductive graphite-coated PS particles also undergo charge transfer, but accumulate substantially less charge, consistent with the distinct charging behaviour expected for conductive materials.
\end{abstract}

\maketitle


\section{Introduction}
\label{Intro}

Triboelectric charging of insulating particles is a ubiquitous yet poorly understood phenomenon \cite{lacks_long-standing_2019}, arising in systems ranging from atmospheric aerosols to granular flows. Despite its prevalence, the mechanisms governing charge transfer during particle–particle collisions remain unresolved, in part due to the difficulty of directly measuring charge exchange at the level of individual interactions.

Understanding the physical processes underlying triboelectric charging between particles has a range of applications throughout the chemical industry and the natural world. The triboelectric charging of granular materials is responsible for electrostatic ignition risks in industry \cite{grosshans_unmasking_2024, ohara_scaling_2025}, affects the lifetimes of aerosols in the atmosphere \cite{mendezharper_lifetime_2022}, electrifies volcanic plumes and Martian dust devils \cite{houghton_triboelectric_2013, reid_lab-based_2024}, and can be harnessed in processes such as electrostatic separation \cite{hotte_electrostatic_2026, labiod_triboelectric_2025}.

For conductive surfaces, charge exchange during contact is comparatively well established and can be described through differences in electronic work functions. When two metals are brought into contact, electrons redistribute between the surfaces until equilibrium is reached, leaving the metal with the higher work function negatively charged after separation. In contrast, the origin of triboelectric charging in insulating materials remains an unresolved question. A range of mechanisms have been proposed, including electron and ion transfer, mechanically driven bond cleavage, and redistribution of material at the contact interface, but the relative importance of each mechanism remains debated \cite{lacks_long-standing_2019, cezan_control_2019}.

Many environmental factors are known to affect the nature and extent of triboelectric charging in insulators, such as surface material \cite{biegaj_surface_2017, ohara_surface_2025, alagha_surface_2026}, relative humidity (RH) \cite{nimvari_investigation_2024}, air-flow turbulence \cite{gemine_influence_2026}, and adventitious (contaminating) carbon \cite{grosjean_adventitious_2026}, among many others. Surface charging is known to exhibit a preferential direction of charge transfer, with one surface consistently tending to acquire either positive or negative charge, even when the surfaces are made as identical as possible \cite{grosjean_asymmetries_2023}. A variety of particle charging models attempt to explain trends observed in experimental particle triboelectric charging, such as the condenser model, which treats a charged particle as analogous to an electrical capacitor, where charge accumulation creates an opposing potential that limits further charge transfer \cite{matsusaka_electrification_2000, matsusaka_electrostatics_2003, grosshans_unifying_2026}. Electrostatic charge on an insulator's surface is also known to form a mosaic of positive and negative sites, which can complicate triboelectric charging \cite{baytekin_mosaic_2011, grosjean_asymmetries_2023}. Recently, it has also been proposed that adsorbed ions of the opposite polarity to a particle's charge are responsible for charge transfer, which can help to explain some unusual triboelectric charging trends, where charge transfer is accelerated at higher initial charges, contrary to the condenser model \cite{jantac_divergent_2026}. 

Bulk powders are of particular interest in triboelectric charging studies due to the impact of charge accumulation on powder handling, transport, and processing \cite{matsusaka_triboelectric_2010}. Particle charge is often measured using techniques such as Faraday cups, which provide the net charge transferred by an enclosed sample. However, these measurements represent an average over many particles, limiting access to individual particle charging behaviour. Efforts have therefore been made to extract additional information from Faraday cup traces \cite{matsusaka_triboelectric_2010, ohara_faraday_2025}. Advances have also been made in particle-tracking voltammetry (PTV) experiments, which can resolve the charge on individual particles falling or in powder flows \cite{xu_spatially_2024, lara_particle_2026}. However, these techniques still lack the precision to accurately resolve the charge transfer over many individual particle collisions; hence, experiments utilising individual particle–particle interactions are required to better understand the fundamentals of triboelectric charging.

There has been a considerable amount of work investigating the charging of individual particles against flat surfaces used to represent particle-particle interactions \cite{matsusaka_electrification_2000, grosjean_single-collision_2023}, including recent work to scale these collisions statistically to more accurately estimate the charge transfer in particle–particle collisions. These find that particle collision statistics often lead to skewed normal distributions in the transferred charge \cite{grosshans_unifying_2026}. However, there has been relatively limited experimental work for individual collisions between isolated particles. Chowdhury \emph{et al.} used streams of air to collide individual particles, which could then be measured using Faraday cups \cite{chowdhury_electrostatic_2021}. This collision approach allows the charge transfer per individual collision to be measured but has a low success rate, with many experiments per successful collision, with the particles landing in their respective cups. This style of approach was built upon by Obukohwo \emph{et al.} by firing a single particle at another held in an acoustic trap \cite{obukohwo_cfd_2023}. This has the advantage of a highly controlled impact velocity and an increased success rate, but still only allows for one collision per pair of particles. Kline \emph{et al.} contacted two levitated particles repeatedly in a Langevin-type acoustic levitator, allowing for repeated measurements of the particles' charge per collision \cite{kline_precision_2020}. They showed that particles charge selectively in one direction with a linear trend in charge per collision at relatively low numbers of collisions. This work aims to give evidence for or against charging models, such as the condenser model, for individual pairs of particles by expanding upon the approach of Kline \emph{et al.} with a novel Faraday cage levitator setup to analyse trends in transferred charge over a greater number of collisions between individual particle pairs.

In this paper, the methodology developed to measure the charge acquired by particles following individual collision events between particle pairs is outlined in Section \ref{Experimental Design}. The developed Faraday cage picoammeters are then calibrated in Section \ref{Calibration}, following the approach of Harrison and Robert \cite{harrison_faraday_2025}. The charging behaviour of repeatedly colliding insulating polystyrene (PS) particles is presented and analysed in Section \ref{Results}. The same system is used to investigate conductive graphite-coated PS particles, providing insights into the nature of triboelectric charging at an individual particle pair level, before the key findings are summarised in Section \ref{Conclusions}.

\section{Methodology}
\label{Methods}
\subsection{Experimental Design}
\label{Experimental Design}
For this experiment, two particles must be brought together and separated repeatedly whilst being levitated such that their charge per collision can be measured. The Levitator employed in this work is the `MultiLev', an enhanced version of the widely used TinyLev \cite{marzo_tinylev_2017}. The MultiLev system uses two arrays of individually controlled transducers to produce a controlled acoustic field that can be manipulated by the transducers' relative phase differences \cite{a_c_barnes_simple_2026}. There are other types of levitator, such as an electrodynamic balance that allows the use of smaller particle sizes, but lacks the fine control of individual particles in three dimensions that the MultiLev can provide \cite{davies_time-resolved_2012}.

\begin{figure}
    \centering
    \includegraphics[width=0.95\linewidth]{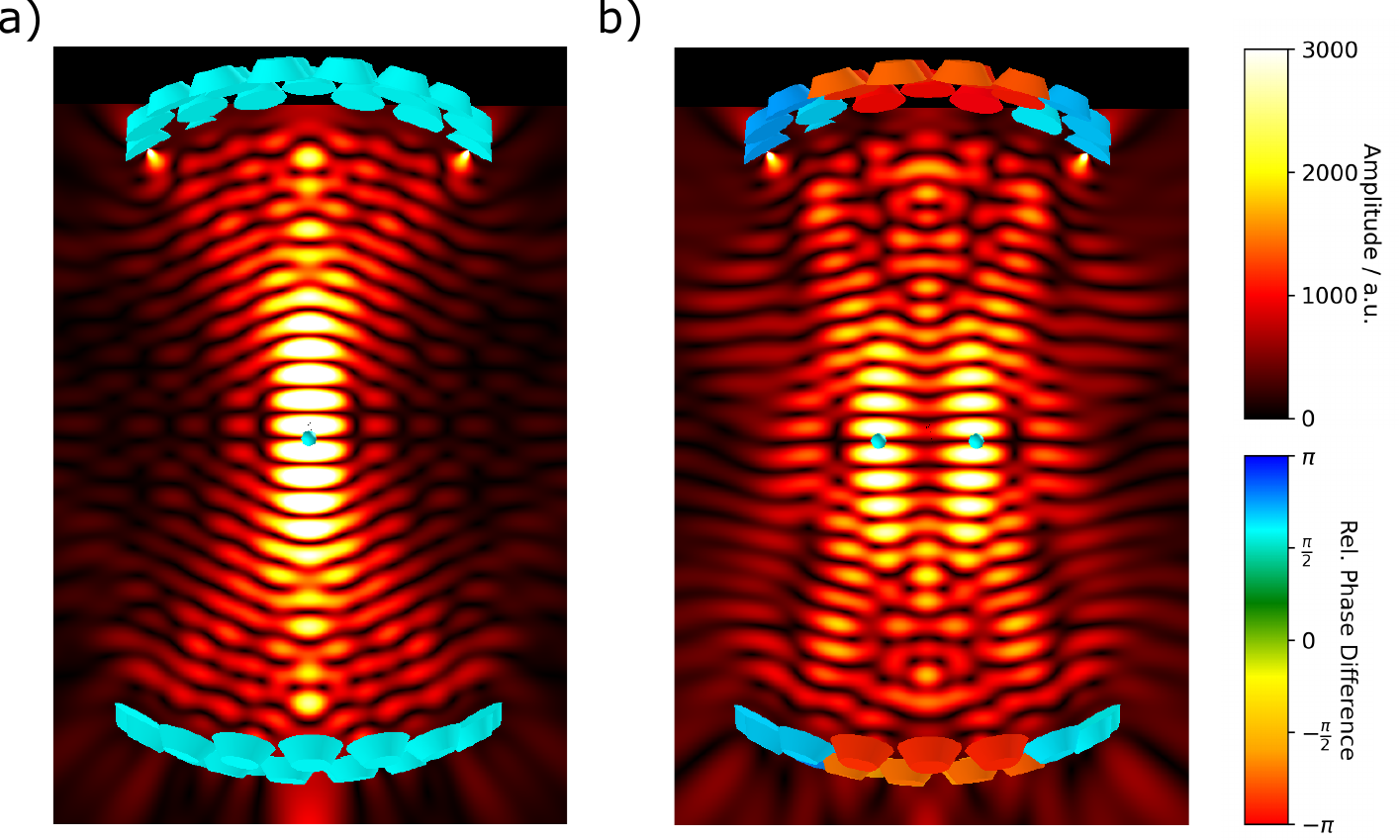}
    \caption[Ultraino acoustic simulations]{Simulations of the acoustic field and pressure and the relative phase differences of transducers required to produce the field for an a) single and b) split trap, where the particles are held together and separate, respectively, produced using Ultraino \cite{marzo_ultraino_2018}. The nodes for levitating PS beads are shown in blue. The coloured scale bars shown apply to both figures, showing the acoustic field amplitude in the XY (vertical and horizontal) plane with the positions of the transducer arrays overlain, coloured by the relative phase difference required to produce the acoustic field.}
    \label{fig:acoustic_sim}
\end{figure}

To calculate the desired acoustic field for the separation and collision of particles, the required phase differences between transducers must be calculated. To achieve this, the open-source Ultraino software developed by Asier Marzo \emph{et. al.} was used \cite{marzo_ultraino_2018}. The two positions required of particles separated and together were simulated, with the results shown in Figure \ref{fig:acoustic_sim}. In the case of a single, central trap as seen in Figure \ref{fig:acoustic_sim}a, all the transducers are in phase, recovering the case of the TinyLev. However, achieving the separated trap shown in Figure \ref{fig:acoustic_sim}b requires the transducers to operate with different relative phases to generate the desired acoustic field. To achieve particle separation and collisions, only the transducer phases for completely separated particle traps and a single trap were used. Many other variations were attempted with smoother simulation steps and other instructions. However, switching between a completely in-phase trap and one where the particles are completely separated, as in Figure \ref{fig:acoustic_sim}, was found to give the most repeatable collision and separation experimentally. This is likely due to the attractive acoustic force between two particles in close proximity that will preferentially push particles together \cite{silva_acoustic_2014}. Therefore, to separate particles effectively, a `hard' switch from the acoustic field with a central trap to one in which there is a peak pressure at the point between the particles is most effective at separating the particles.

To measure the charge on the levitated particles, custom acoustically transparent Faraday cages coupled with picoammeters were designed. The design criteria for the proposed Faraday cages were that they should be sufficiently small to fit within the separation width of the transducer arrays ($<$~100~mm), contain an inner cage electrically isolated from the outer cage, minimise electrical noise such that the signal is greater than the noise, couple to the picoammeter, and remain acoustically transparent.

When choosing the mesh for an acoustically transparent cage, there is a trade-off in the selection of hole size, reducing disturbance to the flow and smaller holes, reducing the scattering of the acoustic waves. To minimise acoustic disturbance, the wire mesh should have a sufficiently high open porosity and low flow resistivity such that viscous losses across the structure remain small compared to the characteristic acoustic impedance of air \cite{jaouen_acoustical_2011}. At the same time, the characteristic aperture scale should remain small relative to the acoustic wavelength to avoid significant scattering \cite{bendali_mathematical_2013}. The minimum spacing between holes ($d$) was shown to be $d < \frac{\lambda}{2}$ by Bendali \emph{et. al.} \cite{bendali_mathematical_2013}. As $\lambda~=~8.65$ mm for 40~kHz acoustic waves at 25 \textdegree C, this means the hole sizes of the mesh should be less than 4.3~mm. For the lower limit of hole size, there should be a large enough open area that the mesh has ``high porosity'', to maintain low flow resistivity. Previous work by Jaouen and B\'ecot used a high porosity material that had an open area of 72\% \cite{jaouen_acoustical_2011}. For this work, a thin galvanised steel square mesh with a wire diameter of 0.2~mm was used with a spacing of 1.5~mm. This means that the open area, as a percentage, is $\frac{(1.5 - 0.2)^2}{1.5^2}\times 100\% \approx 75\%$. This makes the mesh highly porous, whilst also maintaining a spacing of 1.5~mm, less than the required 4.3~mm to avoid significant scattering.

The picoammeter circuits used to measure the current in this setup were based on work carried out by Harrison and Robert, who designed picoammeter circuits for field measurement of small charged objects. This design was suitable to integrate into this setup \cite{harrison_faraday_2025}. Harrison and Robert found that current measurements were optimal for determining particle charge as an alternative to voltage measurement or charge amplification, as it reduced stray capacitances and the requirement for additional bulkier parts such as reed or MOSFET switches \cite{harrison_faraday_2025}.

\begin{figure}
    \centering
    \includegraphics[width=0.8\linewidth]{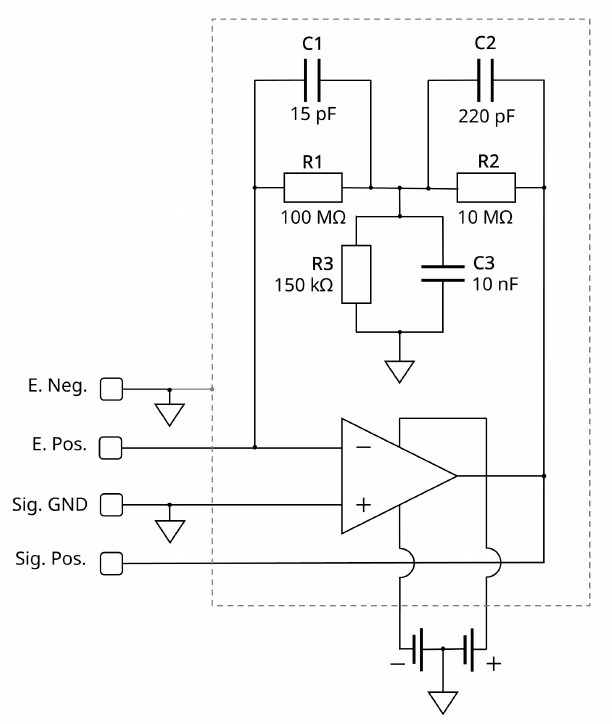}
    \caption{The schematic for the picoammeter circuit with the resistors and capacitors labelled and values shown. The terminals connected to the positive and negative electrodes and the output signal positive and negative are labelled as E. Pos., E. Neg., Sig. Pos., and Sig. GND, respectively. The operational amplifier is an LMC6042; the battery supply is 6~V, with the floating ground shown held at the midpoint of the battery supply. The floating ground is also tied to the shielding casing, represented by the grey dashed line. This circuit is adapted from work by Harrison and Robert \cite{harrison_faraday_2025}.}
    \label{fig:picoammeter_circuit}
\end{figure}


In this circuit, we calculate the effective feedback resistance ($R_f$) using the resistances from R1, R2 and R3 ($R_1 = 100 ~\text{M}\Omega \pm 1\%$, $R_2 = 10~\text{M}\Omega \pm 1\%$, and $R_3 = 150~\text{k}\Omega \pm 0.1\%$, respectively) as  \cite{harrison_faraday_2025}: $R_f = R_1\left(1+\frac{R_2}{R_3}\right)$. These resistor values yield a theoretical value of $R_f = (6.77 \pm 0.10)~\text{G}\Omega$.

\begin{figure*}
    \centering
    \includegraphics[width=0.8\linewidth]{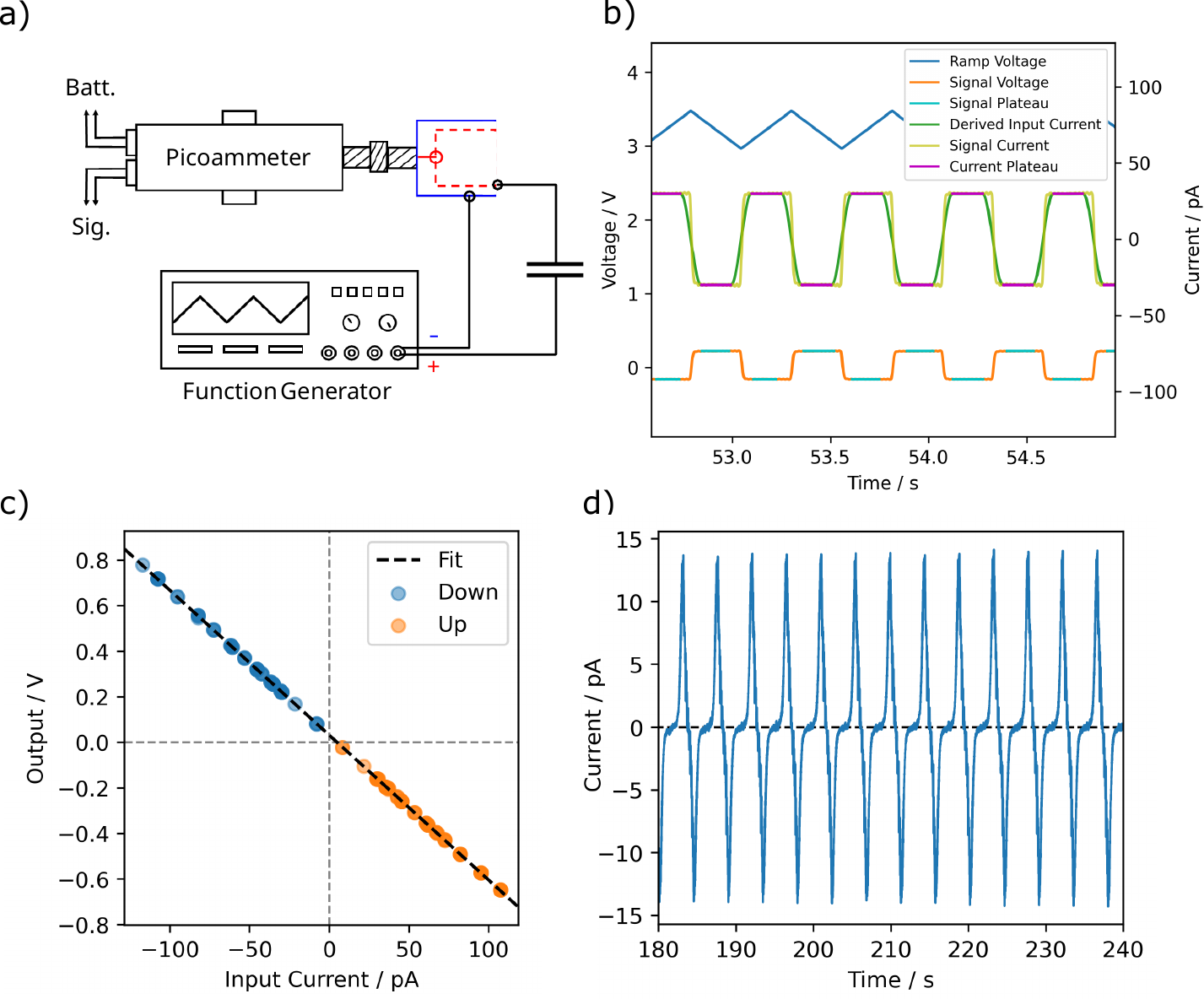}
    \caption[Voltage ramp sensitivity calibration]{a) The schematic for the voltage ramp sensitivity calibration with the polystyrene capacitor value of 15~pF. b) An example section of the measurements taken for the calibration of picoammeter 1. The derived input current is calculated from the input ramp voltage using $I = C\frac{dV}{dt}$, where $C = 15~\text{pF}$. The signal voltage plateaus and the plateaus in the derived current are calculated and used as the input current and output in c). c) The calibration curve, with linear fitting, for the output voltage at positive (Up) and negative (Down) input currents for picoammeter 1 [$\text{fit slope} = - (6.398~\pm~0.0013)~\text{V~nA}^{-1}\text{, y-intercept} = (30.61~\pm~0.09)~\text{mV}$]. d) The picoammeter current from the calibration showing a repeatable current as the particle enters and leaves the cage.}
    \label{fig:calibration}
\end{figure*}

For this device, the time response ($\tau$) is required to be much shorter than the pulse measured. Here, particle collisions and separation occur over a duration of around 60~ms. The $\tau$ of the picoammeter will be at least slower than the RC time constant of the first branch of the T-network. This minimum RC time constant can be found from R1 and C1 (where the capacitance of C1 is $C_1 = 15~\text{pF} \pm 5\%$ as: $R_1 \times C_1 = (1.50 \pm 0.08)~\text{ms}$. However, in the real circuit, the other branches of the T-network, the op-amp bandwidth, and stray capacitances will all affect the time constant. An experimental value for $\tau$ was determined by measuring the picoammeter response to a ramped voltage triangular waveform through a 15~pF capacitor as shown in Figure \ref{fig:calibration}b. The signal current ($I(t)$) over time ($t$) is fit with the function:
\begin{equation}
I(t)=
\begin{cases}
I_0, & t < t_0 \\
I_0 + I_\infty\left(1 - e^{-(t - t_0)/\tau}\right), & t \geq t_0
\end{cases}
\label{eq:exp_transition_fit}
\end{equation}
with the time of the input current step change ($t_0$), the average current before $t_0$ ($I_0$), and the average current long after $t_0$ ($I_\infty$). Averages were required in the calculation of $I_0$ and $I_\infty$ due to the 50~Hz mains noise observed in the signal current. In this calibration, there has been no post-processing to remove this noise to show the signal-to-noise ratio. However, in the remaining results described, this is removed with a 50~Hz passive notch filter of width 1.7~Hz. Overall, this fitting found the characteristic response time for the picoammeter to be $\tau = 7.5~\text{ms}$. After four time constants, the signal should be 98\% of the equilibrated value, assuming a standard RC decay of: $V(t) = V_0\left(e^{-\tau/RC}\right)$. This is characteristic of a device responding rapidly to a step change \cite{harrison_r_giles_meteorological_2015} and shows the constructed picoammeter is comfortably capable of time resolutions around 30~ms (33~Hz), which is more than short enough to capture the induced currents from particles entering the Faraday cages in this work.

\subsection{Picoammeter Calibration}
\label{Calibration}

To convert the picoammeter output voltage into a current measurement and to evaluate the reliability, accuracy, and precision of this part of the system, calibration of the picoammeters with their attached Faraday cages was required. The calibration methodology followed that established by Harrison and Robert \cite{harrison_faraday_2025}. Two complementary calibration procedures were performed for sensitivity and repeatability, respectively. The first sensitivity calibration characterised the voltage response of the picoammeter over a range of known applied currents, generated by applying ramped voltages across a capacitor to produce controlled capacitive currents. The second repeatability calibration assessed the picoammeter response by measuring the signal generated from a repeated physical charging event, in this case, repeatedly dipping a \ac{PS} bead into the Faraday cage.

The first calibration of the picoammeter requires a range of known currents to be applied across the positive and negative electrodes as shown in Figure \ref{Calibration}a. Currents were generated with a \href{https://manualmachine.com/thurlbythandarinstruments/tg230/4644809-user-manual/}{Thurlby Thandar TG230} function generator and a 15~pF polystyrene capacitor. The DC leakage was minimised by the high resistance of the dielectric in the capacitor and by using a DC offset in the ramped voltage waveform that is equal to the midpoint of the battery supply (floating ground at 3~V) \cite{harrison_faraday_2025}. The input signal was a triangular voltage ramp. As the gradient $\frac{dV}{dt}$ remained constant during each half-cycle, the applied current $I = C\frac{dV}{dt}$ had a known constant magnitude. The reversal of the ramp gradient at each turning point caused the current polarity to switch, producing a square-wave current waveform. The frequency and $V_{p-p}$ amplitude were varied to produce a range of applied currents. Both the applied voltage from the function generator and the output voltage from the picoammeter were recorded using a \href{https://www.ni.com/docs/en-US/bundle/usb-6211-specs/page/specs.html}{USB-6210} Data Acquisition (DAQ) device from National Instruments.

The gradient of Figure \ref{Calibration}b, gives $R_f$ from the output voltage ($V_{out}$) and input current ($I_{in}$) as $R_f = -I_{in}V_{out}$ for an ideal inverting op-amp \cite{harrison_r_giles_meteorological_2015}, with the value of $(6.398~\pm~0.0013)~\text{G}\Omega$, which agrees within a few standard deviations of the calculated value of $R_f$ of $(6.77 \pm 0.10)~\text{G}\Omega$. We can also consider the op-amp's non-ideal behaviour of bias current ($I_b$), leakage current ($I_L$) and offset voltage ($V_{OS}$) \cite{harrison_multimode_2000}. This gives an output voltage of $V_{out} = -\left(I + I_b + I_L\right)R_f + V_{OS}$, which can be rearranged to give the functional form of the calibration curve:
\begin{equation}
V_{out} = -R_f I_{in} + \left(V_{OS} - R_f\left(I_b + I_L\right)\right)\text{,}
\label{eq:calibration}
\end{equation}
maintaining the gradient as $R_f$ and making the y-intercept: $V_{OS} - R_f\left(I_b + I_L\right)$. The values obtained in Figure \ref{Calibration} determine the picoammeter's sensitivity and can be used to calculate the input current from the output voltage of the picoammeter, taking into account the leakage and bias current and offset voltage. This process was repeated for the second picoammeter, yielding an $R_f$ value of $(6.365~\pm~0.0023)~\text{G}\Omega$, a gradient of $- (6.365~\pm~0.0023)~\text{V~nA}^{-1}$, which is the picoammeter's sensitivity, and a y-intercept of $(31.95~\pm~0.15)~\text{mV}$.

The second calibration was carried out to test the repeatability of the current and, hence, charge measurement. From Figure \ref{Calibration}b, the current shown (calculated from the picoammeter output voltage using Equation \ref{eq:calibration}) can be converted to the charge on the PS bead through an integration of each peak. This calculation for each peak yielded a mean of: $8.11~\pm~0.11~\text{pC}$ (variance $=~0.012$), giving one standard deviation. This is an error of $1.4\%$.

\begin{figure}
\includegraphics[width=1.0\linewidth]{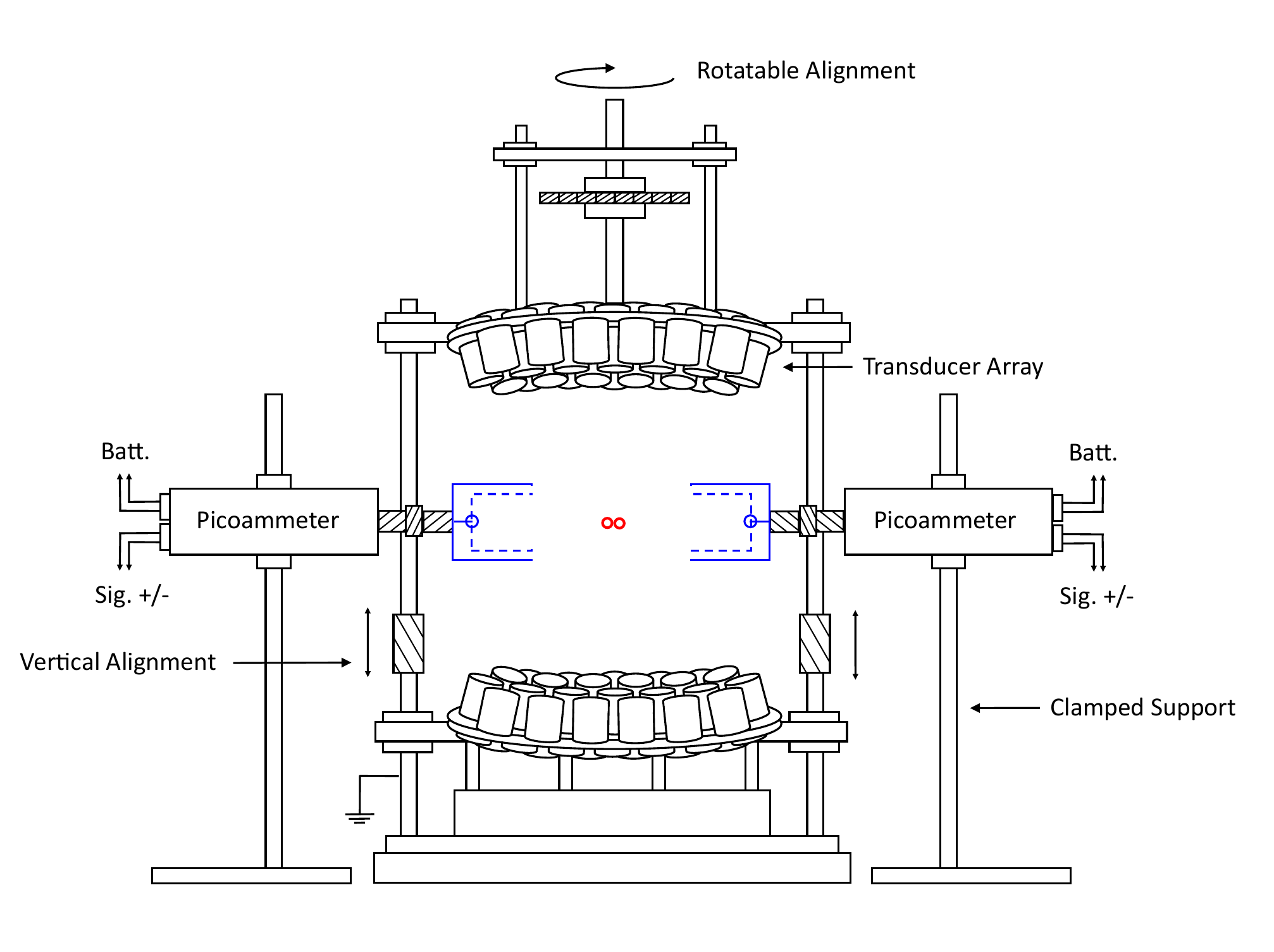}
\caption{The schematic for the MultiLev acoustic levitator, levitating two PS particles (shown in red), combined with acoustically transparent Faraday pails (shown in blue) and picoammeters.}
\label{fig:levitator}
\end{figure}

To combine the acoustic levitator and Faraday cage picoammeters in the physical setup, clamp stands were used such that the orientation and position of the Faraday cages could be easily adjusted. The full experimental setup can be seen in Figure \ref{fig:levitator}. To minimise electrical noise in the picoammeters, any conductive parts of the apparatus were grounded, such as the clamped supports and the alignment frame of the acoustic levitator. As the picoammeter casings were required to be at a floating ground rather than earthed, the clamp jaws were insulating such that the picoammeter casings remained electrically isolated from the common ground.

For the logging of the data from the Faraday cage picoammeters, a \href{https://www.ni.com/docs/en-US/bundle/usb-6211-specs/page/specs.html}{USB-6210} Data Acquisition (DAQ) device from National Instruments was used, which was processed with \href{https://www.ni.com/en/support/downloads/software-products/download.labview.html}{LabVIEW$^{\circledR}$ 2024-Q1} before post-processing and plotting with \href{https://www.python.org/}{Python 3.14.5}, in line with previous work \cite{ohara_faraday_2025}. The DAQ was also connected to a \href{https://datasheet.octopart.com/386-Adafruit-Industries-datasheet-81453130.pdf}{DHT11 humidity and temperature sensor} via an \href{https://docs.arduino.cc/resources/datasheets/A000067-datasheet.pdf}{Arduino Mega 2560} to record the environmental conditions of each drop. The temperature ($T$) and relative humidity (RH) were not controlled directly but were recorded for reference at the time of each drop, with typical values of $T = (23.2 \pm 0.5)~\text{°C}$ and $RH = (44 \pm 5)\%$, throughout the measurement and are reported in the supplementary data files.

\section{Results and Discussion}
\label{Results}

\subsection{Insulating Particles}

The acoustic trap remains in stable operation with two PS particles for up to 5 minutes before instabilities in the air remove a particle from the trap. Following successful measurements of repeated contact and separation between two PS particles, the raw voltage recorded by the picoammeters is converted to current using the sensitivity determined in Section \ref{Calibration}.

\begin{figure}
    \centering
    \includegraphics[width=1.0\linewidth]{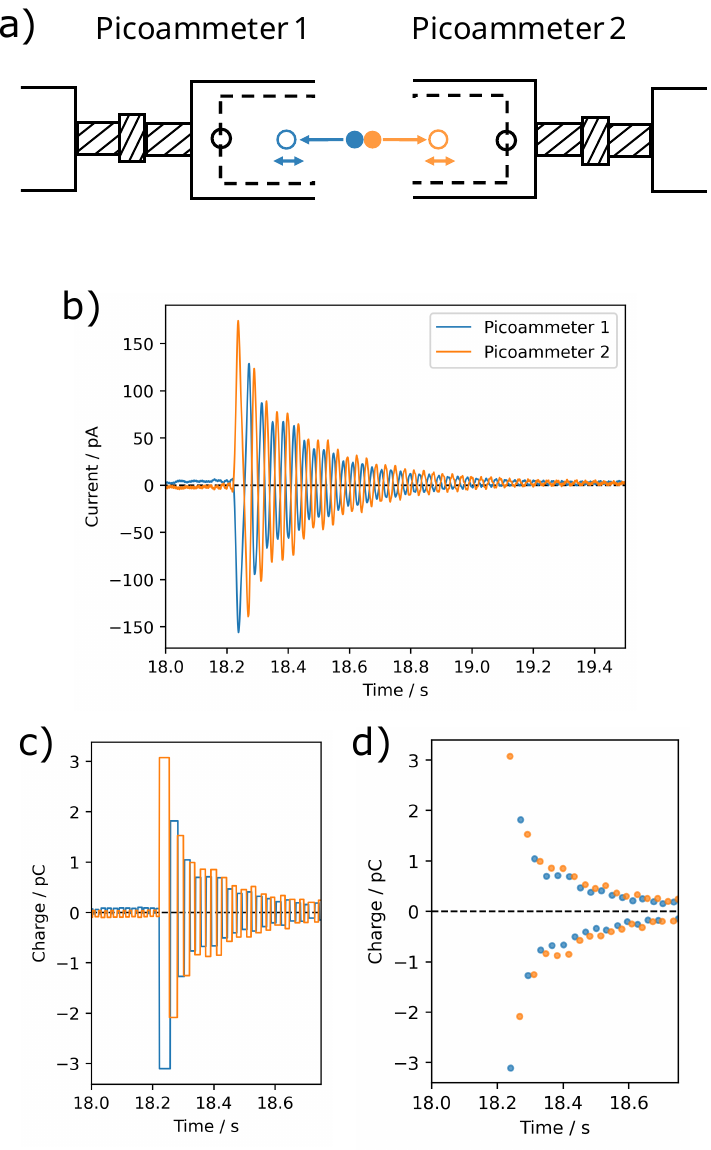}
    \caption[Individual signal processing]{The processing of Faraday cage current signals from repeated collision and separation of two particles in an acoustic field. a) A diagram demonstrating the movement of particles in the acoustic trap separating into the Faraday cages connected to picoammeter 1 and 2, respectively. Their separate positions are marked by filled blue and orange circles before separating to the updated trap positions indicated by the hollow circles, about which damped oscillations occur. b) The output current from both picoammeters shows an individual separation and damped oscillation event. The legend here applies to the following plots. c) The integration of the current with respect to time to get charge, shown as a constant value over each iteration. d) Charge measurements represented as individual points at the middle of their integrated periods.}
    \label{fig:individual_signal}
\end{figure}

The individual separation events give rise to a current signal, as seen in Figure \ref{fig:individual_signal}b. Individual traces show a signal as the trap updates and particles move to their new positions, initially overshooting as they pass the updated trap position, then oscillating about their new location. This can be approximated as a damped harmonic oscillator if the acoustic radiation force ($F_a$) can be assumed to be linearly dependent on its displacement from equilibrium. If $F_a$ arising from the gradient of the Gor'kov potential is approximated as $F_a \approx -k_{ac}x$, where $x$ is the displacement of the particle from its equilibrium position and $k_{ac}$ is the effective stiffness of the acoustic trap. The particle dynamics can then be modelled as: $m\ddot{x} + c\dot{x} + k_{ac}x = 0$, where $m$ is the particle mass and $c$ represents the effective viscous drag coefficient in air. As such, a section of Figure \ref{fig:individual_signal}b is fit in the appendix Figure \ref{fig:Appendix}, yielding frequencies of oscillation ($f$) around $30~\text{Hz}$. This frequency, for a damped harmonic oscillator, will vary with: $f = \frac{1}{2\pi}\sqrt{\frac{k_{ac}}{m}}$. The value of $k_{ac}$ and hence the frequency of oscillation will depend on the amplitude at which the transducers are driven, amongst other factors such as the alignment of the acoustic field and transducer phases.

The frequency of around $30~\text{Hz}$ is in agreement with the high frame rate footage captured using the TinyLev when the trap position is updated, shown in Appendix Figure \ref{fig:individual_signal}, when the particle oscillates about an updated trap position (the full video is included in the supplementary information). It is unlikely to be due to noise, such as from mains power, that would be expected at $50~\text{Hz}$, or one of its harmonics. Additionally, these oscillations are not observed in the previous calibration, where the particles are moved by a string rather than an acoustic trap, as seen in Figure \ref{fig:calibration}d.

\begin{figure*}
    \centering
    \includegraphics[width=0.9\linewidth]{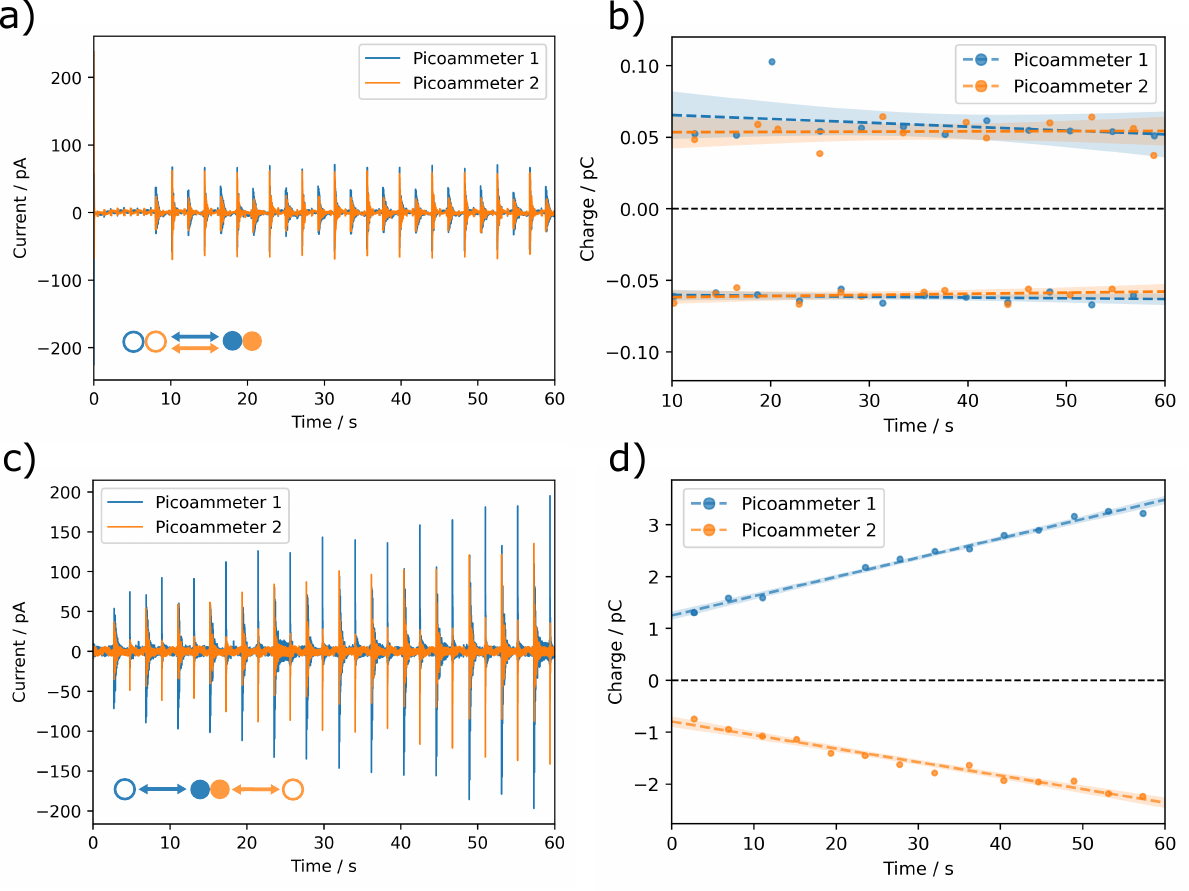}
    \caption[Comparison of Particle Charging Upon Separation]{Measured current traces against time for a) two PS entering and leaving picoammeter 1 without separation and c) two PS particles repeatedly contacting and separating into separated picoammeters 1 and 2, with schematics of particle motion to aid visualisation. The integrated charge values from these current measurements are shown in b) and d), respectively. Linear fits are shown with shaded regions representing the confidence intervals obtained from the statistical uncertainties in the fitted parameters.}
    \label{fig:touching_comparison}
\end{figure*}

To extract accurate particle charge measurements from the individual sections of integrated current measurement shown in Figure~\ref{fig:individual_signal}c and \ref{fig:individual_signal}d, it is important to consider the physical meaning of the induced current signal and how the signals are produced from the particles' motion. The first current peak (shown schematically in Figure~\ref{fig:individual_signal}a) corresponds to the particle moving from the central trap position into the Faraday cage, where the induced current increases as the particle approaches the cage and reaches a maximum near the point of greatest electrostatic induction. As the particle overshoots the updated acoustic trap position, it comes momentarily to rest before accelerating back towards the equilibrium position, causing the induced current to reverse polarity. The subsequent peaks arise from the damped oscillation of the particle about this equilibrium position. Integrating the entire current trace therefore measures the particle moving from the central trap to its final equilibrium position within the cage. However, the particle charge is more accurately represented by the first peak alone, which corresponds to the particle moving from outside the Faraday cage to its maximum displacement within the cage before reversing direction. Including the subsequent oscillations would require differencing positive and negative contributions, increasing the uncertainty without providing additional information about the particle charge. Therefore, only the first current peak is integrated for the remainder of the analysis presented in this work.

As a control, experiments were conducted in which no separation occurred. For two PS particles of roughly equal diameter ($d_p \approx 2~\text{mm}$), the results of both particles not separating but repeatedly moving into one Faraday cage can be seen in Figure \ref{fig:touching_comparison}a and \ref{fig:touching_comparison}c to remain at a roughly constant total charge, within variation. This is as expected, because the two particles remaining together moving into one Faraday cage should not change in net charge, unless other charging/discharging processes were occurring. Although charge may be transferred between the particles, it would not be measured separately as the particles remain together, resulting in a much lower net charge than the particles measured separately.

It is also important to note that the Faraday cages will also be measuring some of the charge of the further particle moving away from it, as this motion also induces a current opposing the current induced by the closer particle. This is due to the size of the cages being comparable to the distance moved by the particle. In an ideal system, the distance moved by the particle would be much larger, such that upon measurement, the more distant particle would not influence the opposing picoammeter measurement. However, under these smaller cage separations, the particle moving away from the Faraday cage produces nearly as strong a signal, although of opposite polarity, as the particle moving into the cage. This can be seen by the response of picoammeter 2 being nearly as large as picoammeter 1 in Figure \ref{fig:touching_comparison}b, despite the particles both moving into picoammeter 1. The charge of one particle cannot be completely separated from an opposing charge in the other at low Faraday cage separations. The Faraday cages were moved to further separations, where the closer particle will dominate the measured charge in the remaining measurements. The cage separation is limited by the particle separation of 20~mm.

In the case of particles repeatedly separating and colliding, seen in Figure \ref{fig:touching_comparison}c and \ref{fig:touching_comparison}d, charging occurs upon collision, leading to an increase in charge over time. Here, after 60~s (15 collisions), the charge of the particles appears to increase linearly with time, yielding $R^2$ values for the positive and negative fits of 0.99 and 0.96, respectively. The gradients for the fittings of $0.074 \pm 0.002$ and $-0.052 \pm 0.003$~pC per collision don't quite overlap within 3 confidence intervals, showing they somewhat disagree. This apparent inconsistency in the conservation of charge most likely arises from asymmetries in the amount of charge detected by the opposing Faraday cage picoammeters at low separations. This asymmetry also led to a slight disparity between the collision and separation signals. The larger currents were produced from particle recombinations (collisions) due to the faster speed as a result of the stiffer acoustic trap with a single focus, rather than the dual trap. Therefore, Figure \ref{fig:touching_comparison}d only shows the larger values produced by particle recombinations, whereas in Figure \ref{fig:touching_comparison}b signals are included from both the particles leaving and entering the cup, as there is no particle separation, so the symmetry is not broken.

\begin{figure}
    \centering
    \includegraphics[width=1.0\linewidth]{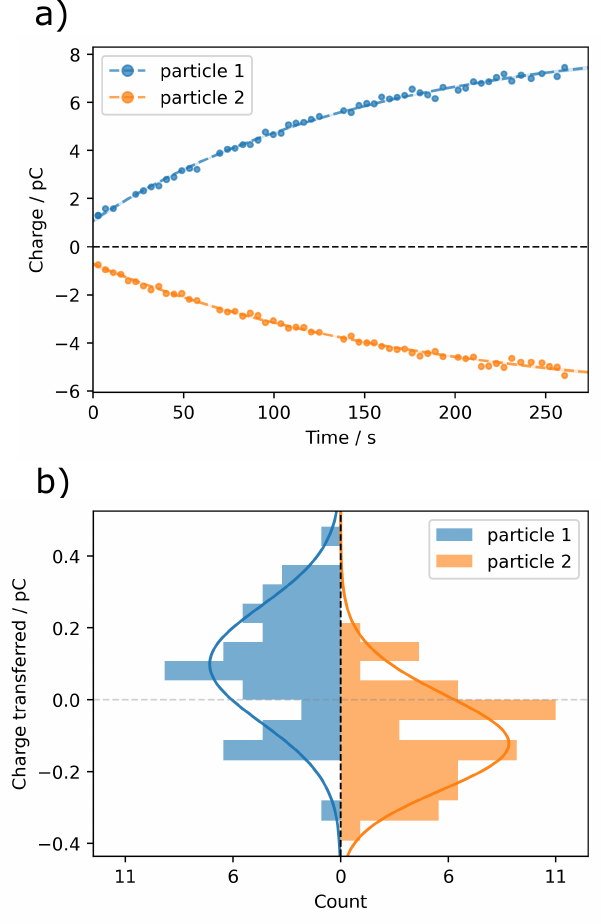}
    \caption[Insulator collisions with condenser model fit]{a) The charge measurements for two PS particles repeatedly contacting in the acoustic trap. Both sets of measurements have been fit with the condenser model of particle charging. b) The charge transferred per collision between the same pair of particles. The distributions for particles 1 and 2 are shown, each fit with an overlain skewed normal distribution, the values of which, along with the condenser model fitting parameters, are tabulated in Table \ref{tab:skew_and_condenser_fits}.}
    \label{fig:insulating_particles}
\end{figure}

We then built upon previous work by extending these measurements over longer time periods, and hence greater numbers of collisions. The linear trend observed with few collisions ($<$~15) in Figure \ref{fig:touching_comparison}d, agrees with previous work by Kline \emph{et al.} but deviates at greater collision numbers ($>$~45), as seen in Figure \ref{fig:insulating_particles}. This new trend supports the condenser model of particle charging, which predicts charging between contacted surfaces until a saturation charge is reached ($Q_{sat}$), analogously to that of an electrical capacitor \cite{matsusaka_electrostatics_2003, grosshans_unifying_2026}. The equation fit for the particle charge ($Q$) as a function of time ($t$) has the form:
\begin{equation}
Q(t)=Q_{sat}\left(1-e^{-\frac{t-t_0}{\tau}}\right),
\end{equation}
where $\tau$ is the characteristic time and $\tau_0$ is the initial time. The fit parameters found are shown in Table \ref{tab:skew_and_condenser_fits}, which show high $R^2$ values ($>0.99$) and low parameter uncertainty ($<2\%$), indicating good agreement with the condenser equation.

\begin{table}[b]
    \centering
    \caption{The condenser model charging fitting parameters and the skewed normal distribution parameters for the contact charging of two acoustically levitated PS particles.}
    \label{tab:skew_and_condenser_fits}
    \begin{ruledtabular}
    \begin{tabular}{lcc}
    Parameter & Particle 1 & Particle 2 \\
    \hline
    \multicolumn{3}{c}{Condenser model charging fits} \\
    $Q_{sat}$ / pC & $8.75 \pm 0.16$ & $6.52 \pm 0.20$ \\
    $\tau$ / s & $154 \pm 6.9$ & $183 \pm 12$ \\
    $t_0$ / s & $-20.41$ & $-21.34$ \\
    $R^2$ & $0.996$ & $0.993$ \\
    \hline
    \multicolumn{3}{c}{Skewed normal distribution fits} \\
    Shape ($\alpha$) & $-0.40 \pm 3.65\times10^{7}$ & $1.21 \pm 3.79\times10^{6}$ \\
    Location ($\xi$) / pC & $0.148 \pm 0.157$ & $-0.213 \pm 0.114$ \\
    Scale ($\omega$) / pC & $0.170 \pm 0.046$ & $0.173 \pm 0.041$ \\
    $R^2$ & $0.645$ & $0.656$ \\
    \end{tabular}
    \end{ruledtabular}
\end{table}

It is important to remember that this fitting is a trend in the charging behaviour, and the charge transferred per collision is not deterministic. This stochastic nature of the triboelectric charging can be observed in Figure \ref{fig:insulating_particles}a, as despite the overall fit being described by the condenser model, individual collisions remain stochastic. The charge per collision from the same dataset as Figure \ref{fig:insulating_particles}a can be seen distributed in Figure \ref{fig:insulating_particles}b. The distribution of charge transferred per collision resembles a skewed normal distribution in line with previous work \cite{grosshans_unifying_2026}. The skewed-normal distributions used were of the Azzalini form and follow the equation \cite{azzalini_class_1985}:
\begin{equation}
f(x)=\frac{2}{\omega}\,\phi\!\left(\frac{x-\xi}{\omega}\right)\,
\Phi\!\left(\alpha\frac{x-\xi}{\omega}\right),
\end{equation}
given:
\begin{equation}
    \phi(z)=\frac{1}{\sqrt{2\pi}}e^{-z^2/2},
    \quad
    \Phi(z)=\int_{-\infty}^{z}\phi(t)\,dt,
\end{equation}
where $\xi$ is the location parameter, $\omega>0$ is the scale parameter, and $\alpha$ is the shape (skewness) parameter controlling the degree and direction of asymmetry. The fitting parameters used in Figure \ref{fig:insulating_particles}b can be seen in Table \ref{tab:skew_and_condenser_fits}. The fits have relatively low $R^2$ values and have large variability in shape due to only 55 collisions occurring between the particles. The shape parameters ($\alpha$) are not statistically dissimilar to zero, so the distributions could also be described by normal distributions. However, here we have used skewed normal distributions to be consistent with previous work, which has taken many more collisions ($>$~200) of different PS particles with surfaces \cite{grosshans_unifying_2026}. Here, the location parameters, which are analogous to the means of unskewed normal distributions, are of opposite polarity and agree within one standard deviation, meaning that charge is conserved for the normal distribution case, where the location parameter is the mean charge transferred.

\subsection{Conductive Particles}

To contrast the charging occurring between the insulating PS particles and conductive particles, a conductive graphite spray coating was applied to the PS particles before being repeatedly contacted in the levitation setup. This coating gives the PS particles a conductive surface layer, whilst maintaining a low density. It is important to note that a conductively coated particle may behave differently from a particle that also has a conductive core. For example, the internal conductivity may affect intra-particle relaxation of charge between sites where particle charging may occur. However, as triboelectric charging is a surface phenomenon, we assume here that the particle's surface conductivity will dominate whether its triboelectric charging behaviour acts as an insulating or conductive particle.

\begin{figure}
    \centering
    \includegraphics[width=1.0\linewidth]{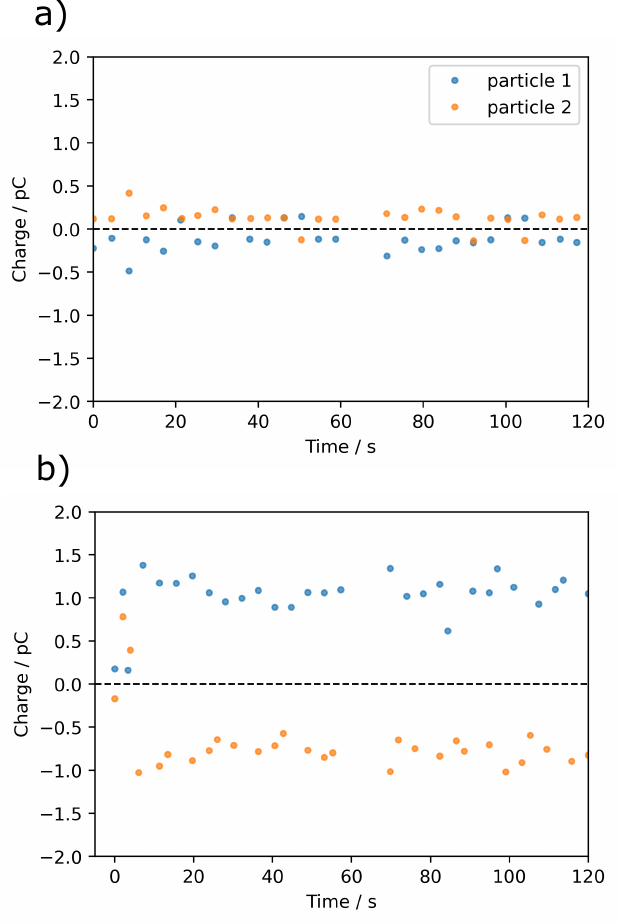}
    \caption[Conductive particle charging]{a) The charge measured on a pair of graphite-coated PS particles repeatedly colliding and separating in the acoustic trap. b) The charge measured on an uncoated PS particle (particle 1) and a graphene-coated PS particle (particle 2) during repeated collisions. The particle labels shown in a) are used for both panels, although different particle pairs were used in each experiment.}
    \label{fig:conductive_particles}
\end{figure}

When two conductive particles are repeatedly contacted, around an order of magnitude less charging occurs, as shown in Figure \ref{fig:conductive_particles}a, compared with the case where one particle is insulating (Figure \ref{fig:conductive_particles}b). The charging is also substantially lower than that observed when both particles are insulating after multiple collisions (Figure \ref{fig:insulating_particles}). It is typical that insulator triboelectric charging is of a lower magnitude than that of conductors. Here, the charging is such that for most collisions, the charge is almost indistinguishable from the background noise. However, for some collisions, a comparatively larger amount of charging occurs before relaxing upon the next collision. This charging behaviour indicates that the sprayed coating may be discontinuous, leaving some isolated patches of the interior particles' insulating surface exposed. This exposed region would charge upon impact; then, upon the next collision, if that area collides with a majority conductive area, this charge can be transferred and dissipated.

The lack of an increasing trend in particle charge also raises the question of which region of the particle surface is involved in each collision event. Rotation of particles has been observed in TinyLev-type acoustic levitation systems and is generally attributed to acoustic streaming (steady airflow generated by the acoustic field) and acoustic radiation torque (torques arising from asymmetric acoustic forces acting on the particle) \cite{mcelligott_tinylev_2022}. However, in the present setup, only minimal rotational motion was typically observed during operation. If more detailed modelling of the contact mechanics were undertaken, the effect of any rotation would need to be accounted for in a time-resolved description of the particle surface orientation during collisions. The total area involved in the contact will also be affected by the impact speed of the rough surface as the particles deform on impact \cite{jantac_triboelectric_2025}. As the driving voltage was held constant during the experiments at around $12~V$, the force due to the acoustic field and hence the impact velocity should remain constant for a pair of particles. Future experiments could vary $V_{p-p}$ to deliberately investigate different impact velocities. However, if the collisions do not need to be repeated, then other methods are available to produce a wider range of impact velocities, such as the pneumatic conveying setup combined with acoustic levitation developed by Obukohwo \emph{et al.} \cite{obukohwo_novel_2026}.

When one conductive graphite-coated PS particle and one insulating uncoated PS particle repeatedly contact and separate, as shown in Figure \ref{fig:conductive_particles}b, only a relatively small amount of charge is exchanged, with a rough equilibrium charge being reached after very few collisions. The particles then continue to transfer charge stochastically but remain around this new equilibrium charge level. It is also likely that some graphite is transferred between the two particles and may be responsible for some of the charge transfer. Nevertheless, these results are still valuable because they demonstrate that even partial surface conductivity fundamentally alters the evolution of charge transfer during repeated particle contacts, providing insight into how material conductivity influences triboelectric charging.

\section{Conclusions}
\label{Conclusions}

The acoustic levitation and charge measurement system developed here has enabled the direct, repeated measurement of charge transfer between pairs of particles, for longer times than was previously possible. At low numbers of collisions of insulating particles, our results are consistent with previous work \cite{kline_precision_2020}. At higher collision numbers, we demonstrate that acoustically levitated pairs of insulating particles tend towards a saturation charge consistent with the condenser model, despite the stochastic nature of individual collision events, which can be described by a skewed normal distribution. Conductive particles are also observed to transfer significantly less charge via triboelectric charging within the same system.

However, limitations remain. In the current setup, only low-density materials such as PS and graphene-coated PS can be used, and the measurement of particle charges cannot be fully separated at low cage separations. In future work, a levitator with higher-power transducers and a modified geometry, for example, a dual half-pipe configuration, could enable stable levitation and greater separation of higher-density materials, addressing both of these limitations.

Despite these limitations, this work provides the first direct experimental evidence that repeated collisions between an acoustically levitated pair of identical insulating particles evolve towards the saturation behaviour predicted by the condenser model. This establishes acoustic levitation as a powerful platform for testing models of triboelectric charging at the single-particle level and opens the way to investigating a much wider range of materials and contact conditions.

\section*{Data Availability Statement}
All data associated with this article, including picoammeter measurements, acoustic simulation outputs, and high-speed imaging data, are available through the Materials Data Facility under the GPL-4.0 license \href{https://materialsdatafacility.org/detail/b3740e10-bcc5-454c-a00b-89d74a147230-1.0}{(DOI: 10.18126/DK7G-3K88)}. The source code for the MultiLev control software and the data-processing scripts used to reproduce the figures are available under the GPL-3.0 license \href{https://zenodo.org/doi/10.5281/zenodo.21264875}{(DOI: 10.5281/zenodo.21264875)}.

\begin{acknowledgments}
The authors acknowledge Allegra Skare, Dr James Drewitt, and Harry Hannaford from the University of Bristol School of Physics for their instrumental support with the MultiLev apparatus. The authors also thank Prof. Giles Harrison from the University of Reading School of Mathematical, Physical and Computational Sciences for constructive discussions and input that enabled the adaptation of the picoammeter design. We acknowledge David Reid and George Burns from the University of Bristol School of Civil, Aerospace and Design Engineering Laboratory for their technical and electronics expertise in assisting with the construction of the Faraday-cage picoammeters. Finally, we thank Dr Daniel Mitchard and Meirion Hills from the University of Cardiff School of Engineering for providing the high-frame-rate camera and assistance with its operation.
\end{acknowledgments}

\section*{Appendix}

This appendix presents schlieren imaging snapshots of a PS particle oscillating in an acoustic field, analysis of the oscillations using a damped harmonic oscillator model, and simulations of the picoammeter circuit response to ramped voltage inputs.

\begin{figure}
    \centering
    \includegraphics[width=1.0\linewidth]{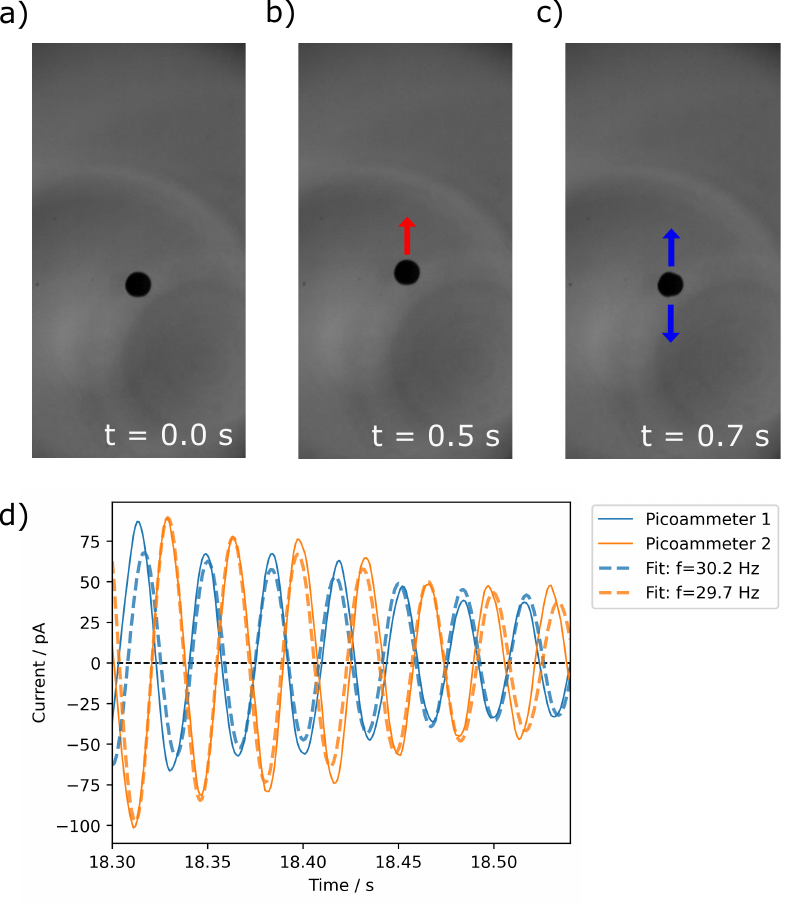}
    \caption[Schlieren images of levitation]{Schlieren images of a PS particle levitated within the TinyLev apparatus. Snapshots are shown at a) the initial position, b) after gradually moving the particle up, and c) where the particle is oscillating back around its original position. The full video is available in the supplementary information. For the still images, the arrows indicate the particle motion. d) The current traces from a PS separation event are fit with a damped harmonic oscillation function, with the fitted oscillation frequencies ($f$) shown. The parameter values for the picoammeter 1 and 2 fittings are laid out in Table \ref{tab:damped_harmonic_fit}.}
    \label{fig:Appendix}
\end{figure}

The damped harmonic oscillator model used to fit the picoammeter current traces, shown in Figure \ref{fig:Appendix}, is:
\begin{equation}
    I(t) = A e^{-\gamma (t-t_0)} \cos\left(2\pi f (t-t_0) + \phi\right) + y_0
\end{equation}
where $t_0 = 18.3~s$ is the start of the fitting region, and the remaining parameters are defined in Table \ref{tab:damped_harmonic_fit}. The $R^2$ values (0.84 and 0.92), Q-factors (35.8 and 22.0), and low parameter uncertainties ($3\%$ to $10\%$) indicate reasonable agreement with the fitting. However, the $\chi^2$ values $>>~1$ indicate that not all features are captured, including those arising from the non-linear dependence of the Gor'kov potential on the distance from equilibrium and other considerations outlined by Morrell \emph{et al.} \cite{morrell_acoustodynamic_2023}.

\begin{table}[b]
    \centering
    \caption{Damped harmonic oscillator parameters used for the fitting in Figure \ref{fig:Appendix}, with the quality of fit statistics shown for picoammeter 1 and 2.}
    \label{tab:damped_harmonic_fit}
    \begin{ruledtabular}
    \begin{tabular}{lcc}
    Parameter & Picoammeter 1 & Picoammeter 2 \\
    \hline
    Number of points ($N$) & $241$ & $241$ \\
    $R^2$ & $0.841$ & $0.916$ \\
    Reduced $\chi^2$ & $248$ & $205$ \\
    \hline
    Amplitude ($A$) & $66.9 \pm 3.3$ & $102.0 \pm 3.4$ \\
    Damping ($\gamma$) / s$^{-1}$ & $2.63 \pm 0.42$ & $4.20 \pm 0.32$ \\
    Frequency ($f$) / Hz & $29.94 \pm 0.07$ & $29.33 \pm 0.05$ \\
    Period ($T$) / s & $0.0334$ & $0.0341$ \\
    Phase ($\phi$) / rad & $3.14 \pm 0.05$ & $0.900 \pm 0.031$ \\
    Offset ($y_0$) & $3.99 \pm 1.02$ & $-0.69 \pm 0.93$ \\
    \hline
    Quality factor ($Q$) & $35.8$ & $22.0$ \\
    \end{tabular}
    \end{ruledtabular}
\end{table}

\begin{figure}
    \centering
    \includegraphics[width=1.0\linewidth]{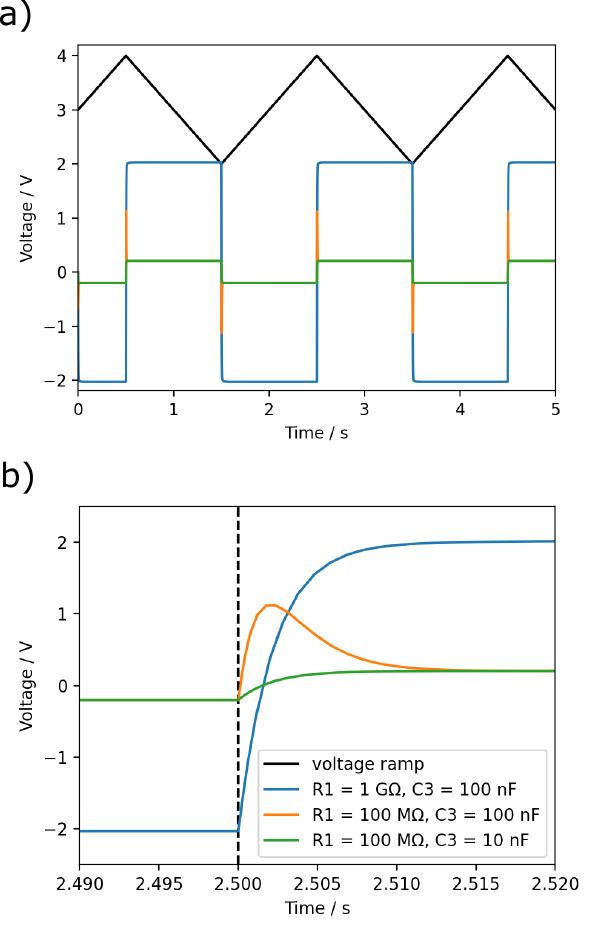}
    \caption[Simulations of the picoammeter time response]{a) Simulations of the picoammeter sensitivity to the ramped voltage used in the calibration. b) A zoomed-in version of the same simulations showing the time response of the resistor (R1) and capacitor (C3) values used, with the legend applying to both plots. The dashed black line indicates the point at which the gradient in the triangular waveform switches, to aid with the visualisation of the time response.}
    \label{fig:LTSpice}
\end{figure}

The LTspice simulations in Figure \ref{fig:LTSpice} show the time response and sensitivity of the picoammeters constructed with varying R1 and C3 values. Corresponding to the resistors and capacitors labelled in Figure \ref{fig:picoammeter_circuit}. The circuit values selected for this work were: R1 $=$ 100 M$\Omega$, C3~$=$~10 nF, as they demonstrated the reduced sensitivity required, with no overshoot of the signal response.

\FloatBarrier

\bibliography{references}

\end{document}